\documentclass[amsmath, amssymb, aps, preprint, bibliography]{revtex4-1}
\usepackage{amssymb, amsmath, bm, graphicx,color}
\usepackage{epstopdf}
\usepackage{array}
\begin{document}

%\preprint{}

\title{Surface-plasmon-enhanced near-field radiative heat transfer between planar surfaces with a thin-film plasmonic coupler}


\author{Mikyung Lim$^{1,\dagger}$, Jaeman Song$^{2,3,\dagger}$, Seung S. Lee$^2$, Jungchul Lee$^{2,3}$, Bong Jae Lee$^{2,3}$}
\email{bongjae.lee@kaist.ac.kr \\}
%\thanks{}
\affiliation{1. Korea Institute of Machinery and Materials, Daejeon 34103, South Korea\\
2. Department of Mechanical Engineering, Korea Advanced Institute of Science and Technology, Daejeon 34141, South Korea\\
3. Center for Extreme Thermal Physics and Manufacturing, Korea Advanced Institute of Science and Technology, Daejeon 34141, South Korea\\
$^\dagger$These authors contributed equally to this work.\\}

%option in \documentclass).\noaffiliation is required (may also be

\date{\today}

\begin{abstract}

In last decade, there have been enormous efforts to experimentally show the near-field enhancement of radiative heat transfer between planar structures. Several recent experiments also have striven to achieve further enhanced heat transfer with the excitation of coupled surface polaritons by introducing nanostructures on both emitter and the receiver; however, these symmetric structures are hardly employed in real-world applications. Here, we demonstrate substantially increased near-field radiative heat transfer between asymmetric structures (i.e., doped Si and SiO$_2$) by using a thin Ti film as a plasmonic coupler. The measured near-field enhancement at vacuum gap of 380 nm is found to be 3.5 times greater than that for the case without the coupler. The enhancement mechanism is thoroughly elucidated for both polarizations and a dimensionless parameter, which can quantify the coupling strength of the surface polaritons at vacuum, is suggested. As a thin film can be readily used in many engineering applications, this study will facilitate the development of the high-performance engineering devices exploiting the near-field thermal radiation.

\end{abstract}
%\pacs{44.40.+a; 78.20.Ci}
%\keywords{}
\maketitle


It is well known that radiative heat transfer between two spatially close media can exceed the blackbody limit via tunneling of evanescent waves, which exist exclusively near surfaces \cite{polder1971theory, mulet2002enhanced, joulain2005surface, basu2010near}. This phenomenon, so called near-field radiative heat transfer, has recently drawn enormous attention because of its tunability using nanostructures \cite{tian2018review} and its potential applications in thermophotovoltaics (TPV) \cite{laroche2006near, basu2007microscale, park2008performance, st2017hot, fiorino2018nanogap, vongsoasup2018effects, inoue2019one}, photonic cooling \cite{zhu2019near}, and thermal diode \cite{fiorino2018thermal, ito2017dynamic}. To develop a high-performance device for those emerging engineering applications, large planar structures separated by sub-micron gap with substantial temperature difference are required. Accordingly, continuous efforts have been made towards measurement of a remarkable heat transfer between planar structures \cite{hargreaves1969anomalous, domoto1970experimental, hu2008near, ottens2011near, kralik2012strong, ito2015parallel, lim2015near, song2016radiative, watjen2016near, bernardi2016radiative, ghashami2018precision, fiorino2018giant,desutter2019near}, mostly with the homogeneous bulk media where radiative heat transfer is readily determined by dielectric function of the medium. 

Recently, a few groups have successfully demonstrated `tunable' near-field radiative heat transfer by introducing a monolayer graphene \cite{yang2018observing, thomas2019electronic, shi2019colossal} and metallo-dielectric multilayers \cite{lim2018tailoring} on both the emitter and the receiver surfaces. These planar nanostructures are compatible with the existing experimental platforms \cite{hu2008near, ottens2011near, kralik2012strong, ito2015parallel, lim2015near, song2016radiative, watjen2016near, bernardi2016radiative, ghashami2018precision, fiorino2018giant,desutter2019near} and are known to change the condition of surface plasmon polaritons supported at the vacuum/emitter and the vacuum/receiver interfaces \cite{biehs2007thermal, francoeur2008near, messina2013graphene, lim2013near, karalis2016squeezing, iizuka2018significant}. Considering that the tunneling of evanescent waves can be notably enhanced by the coupling of surface polaritons at the vacuum/emitter and the vacuum/receiver interfaces \cite{messina2013graphene, lim2013near,karalis2016squeezing}, both intensity and spectral distribution of the near-field radiation can be tuned by introducing nanostructures on each surface \cite{messina2013graphene, lim2013near, karalis2016squeezing, lim2018optimization}. For example, it was shown that by modifying the configuration of metallo-dielectric multilayers (e.g., thickness of each layer or number of unit cells), the surface plasmon polariton (SPP) conditions of the interfaces near vacuum can be tuned, which in turn can lead to the enhanced total radiative heat flux \cite{lim2018tailoring}. On the other hand, the near-field radiative heat transfer can also be greatly increased by placing graphene on the surfaces of intrinsic silicons and consequently making both the emitter and the receiver (i.e., graphene-coated silicons) to support SPPs \cite{yang2018observing}. Because SPPs generated at the graphene layer can be tuned by its chemical potential, a demonstration of an electronic modulation of near-field radiative heat transfer was also reported by applying electronic bias on the graphene layer \cite{thomas2019electronic}. Further, the SPPs of graphene can be coupled with the surface phonon polaritons (SPhPs) of SiO$_2$ substrates, such that the colossal enhancement (i.e., $\sim$ 65 times) over the blackbody limit was achieved with a pair of graphene-coated SiO$_2$ structures \cite{shi2019colossal} by coupling of SPP-SPhPs supported at vacuum/emitter and vacuum/receiver interfaces. The authors also noted that this significant near-field enhancement over the blackbody radiation is greatly suppressed if the symmetry between the emitter and the receiver is broken. In fact, graphene/SiO$_2$-to-SiO$_2$ structure shows smaller heat transfer than that between SiO$_2$ structures or that between graphene-coated SiO$_2$ structures, because the resonant conditions (i.e., surface polariton conditions) of vacuum/emitter and vacuum/receiver interfaces become hardly matched with the asymmetric structures \cite{shi2019colossal}.

Despite all those great advances, researches on tuning of the near-field radiative heat transfer has focused mainly on obtaining a large heat flux by introducing the same materials on both the emitter and the receiver sides (i.e., symmetric configuration) \cite{yang2018observing, thomas2019electronic, shi2019colossal, lim2018tailoring}, which is, however, hard to be achieved in real-world applications \cite{laroche2006near, basu2007microscale, park2008performance, fiorino2018nanogap, vongsoasup2018effects, inoue2019one, zhu2019near, fiorino2018thermal, ito2017dynamic}. For example, the near-field TPV system (one of the most promising near-field radiative heat transfer applications) requires a TPV-cell receiver and a selective emitter for a high performance \cite{laroche2006near, park2008performance, vongsoasup2018effects, lim2018optimization}. Even if the same materials are chosen for both sides, temperature-dependent optical (and/or thermophysical) properties of the materials make the overall system to be asymmetric \cite{fu2006nanoscale}.

\begin{figure}[!t]
\centering\includegraphics[width=0.95\textwidth]{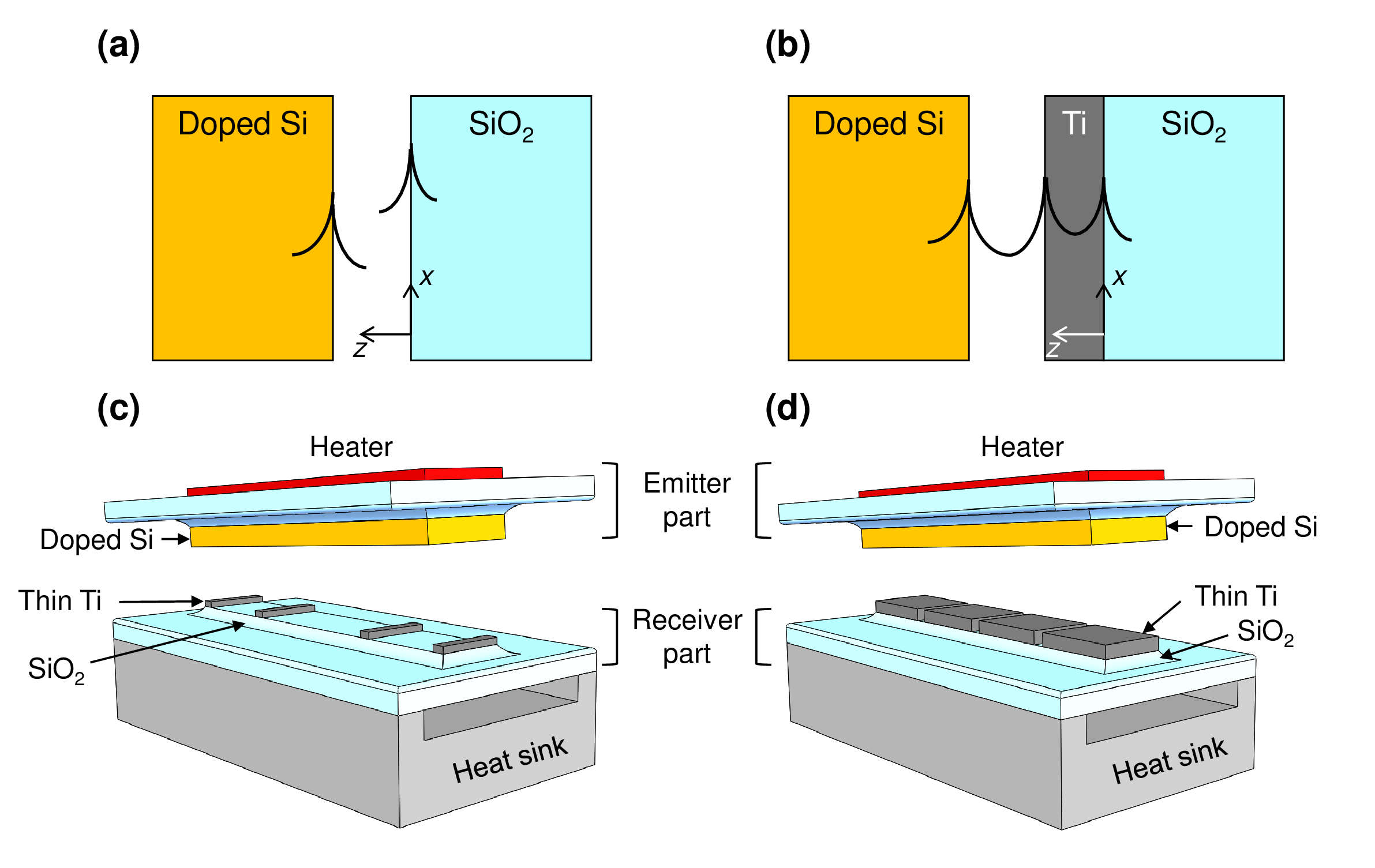}
\caption{Schematics of configurations of \textbf{(a)} doped-Si emitter and bare-SiO$_2$ receiver (i.e., without plasmonic coupler)  and \textbf{(b)} doped-Si emitter and 10-nm-Ti-film-coated SiO$_2$ receiver (i.e., with plasmonic coupler). The conceptual field profiles for SPPs (SPhPs) supported at the interfaces are depicted. Schematics for the emitter part and the receiver part of the microdevices for measuring near-field radiation \textbf{(c)} between doped-Si emitter and bare-SiO$_2$ receiver and \textbf{(d)} between doped-Si emitter and Ti-film-coated-SiO$_2$ receiver.}
\label{Fig:1}
\end{figure}

In this work, we experimentally demonstrate significantly enhanced near-field radiative heat transfer between asymmetric emitter and receiver by introducing a thin metal film as a plasmonic coupler. As shown in Fig.\ \ref{Fig:1}(a), doped Si and SiO$_2$ are used as an emitter and a receiver, respectively.  Although doped Si and SiO$_2$ are well-known to support SPPs and SPhPs, respectively, their resonant frequencies do not overlap, which makes the coupling of surface polaritons of the emitter and the receiver occurring in the limited frequency range. Thus, there would be little synergetic effect between SPPs (associated with doped Si) and SPhPs (associated with SiO$_2$) for enhancing the near-field heat transfer rate. If a thin Ti film is deposited on the SiO$_2$ side [see Fig.\ \ref{Fig:1}(b)], the SPPs generated at vacuum/Ti/SiO$_2$ interfaces can be effectively coupled to SPPs of vacuum/doped Si interface in a wider frequency range, and thus, enhances the net radiative heat transfer. The measured heat flux for these two configurations (i.e., with or without Ti-film plasmonic coupler) are well-matched with the theoretical predictions. Further, for the analysis of enhanced heat transfer via $p$-polarization through coupling of surface polaritons, we suggest a simple dimensionless parameter that is related to the field distribution at the vacuum to quantify the coupling strength of SPPs of the emitter and the receiver.

%%%%%%%%%%
%%%%%%%

To measure the near-field radiation between two media depicted in Figs.\ \ref{Fig:1}(a)-\ref{Fig:1}(b), MEMS-fabricated microdevices and a custom-built three-axis nanopositioner, which was introduced in the previous work \cite{lim2018tailoring}, are employed. As described in Figs.\ 1(c)-1(d), the emitter part of the microdevice is composed of the 800-nm-thick doped-Si layer deposited on one side of a fused-silica substrate and the resistive heater on the opposite side. For the receiver part of the microdevice, two different configurations are employed for comparison: one is bare SiO$_2$ [see Fig.\ \ref{Fig:1}(c)] and the other is 10-nm-thick-Ti-film-coated SiO$_2$ [see Fig.\ \ref{Fig:1}(d)]. The detailed information on the fabrication process, the schematics, and photographs of the emitter and the receiver parts of the microdevice are provided in Supplemental material. 

To precisely estimate the vacuum gap between the emitter and the receiver, four 10-nm-thick Ti electrodes with small area (i.e., 2.6\% of total receiver surface for each electrode) are deposited on the bare SiO$_2$ receiver surface in case of the bare-SiO$_2$ receiver [see Fig.\ \ref{Fig:1}(c)], while four-segmented 10-nm-thick-Ti films themselves are used as electrodes in case of the 10-nm-thick-Ti-film-coated-SiO$_2$ receiver, as can be seen in Fig.\ \ref{Fig:1}(d). Following the previous work \cite{lim2018tailoring}, local vacuum gaps between doped-Si-electrode of the emitter and each of four thin-Ti-film electrodes of the receiver are estimated by sequentially measuring the electrical capacitances between them. Thus, we can fully quantify the parallelism between the emitter and the receiver, as well as take into consideration the effects of bowing and tilting on the radiative heat flux by using Derjaguin-approximated average vacuum gap \cite{derjaguin1956direct}. 

%%%%%%%
\begin{figure}[!t]
\centering\includegraphics[width=0.6\textwidth]{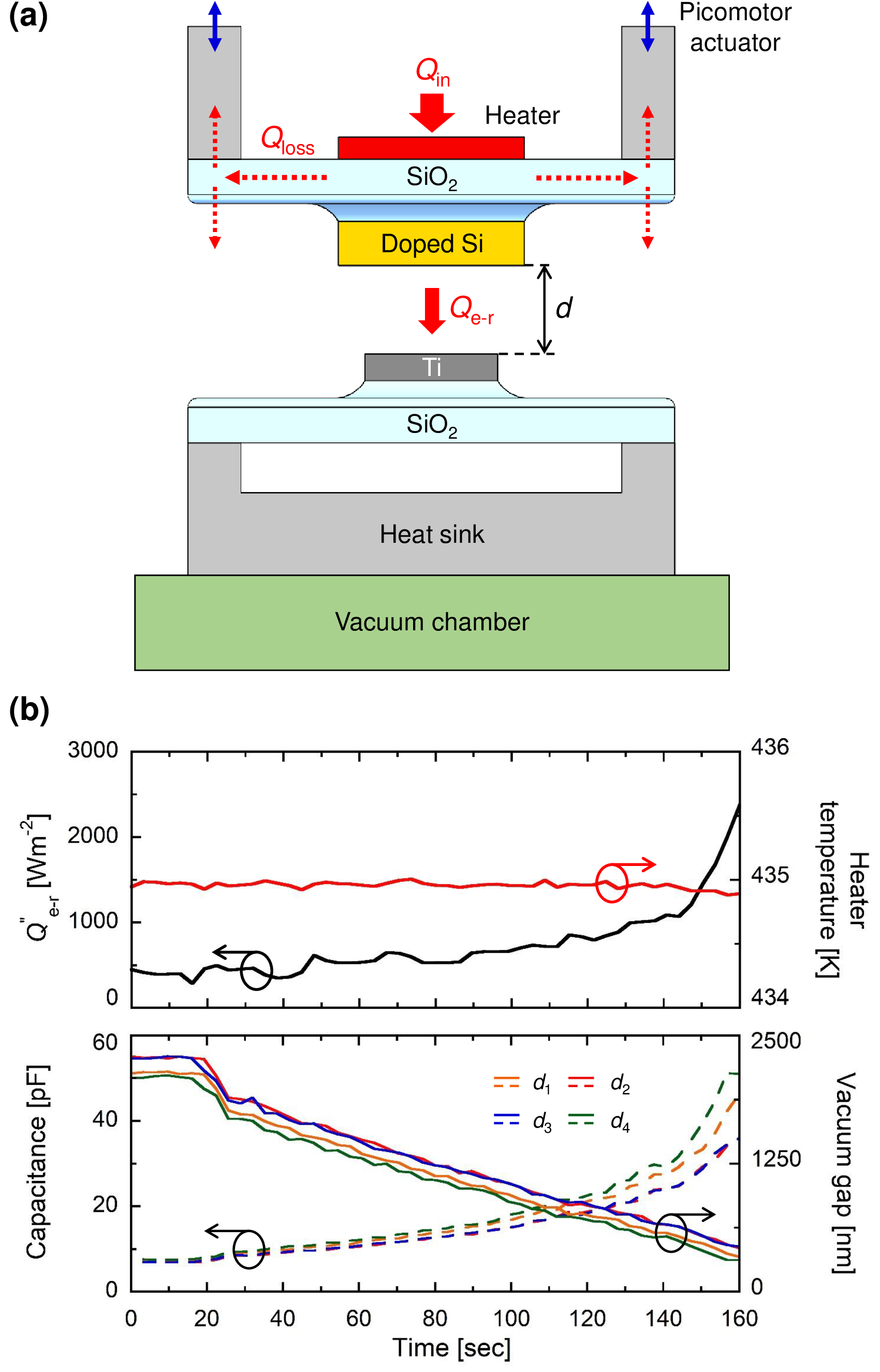}
\caption{\textbf{(a)} Schematic of the cross sectional view of the microdevice attached to the custom-built nanopositioner. The heat flow from the input power of the heater, $Q_{\text{in}}$ into the radiative heat transfer from emitter to receiver, $Q_{\text{e-r}}$ and $Q_{\text{loss}}$ is depicted. \textbf{(b)} Measured change of $Q_{\text{in}}$ to maintain the heater temperature constant while reducing the gap (upper panel) and four measured capacitances and corresponding four local gaps while conducting one cycle of measurement (lower panel).}
\label{Fig:2}
\end{figure}
%%%%%%%

Given that the experimental setup is placed in a high vacuum chamber ($<1\times10^{-3}$ Pa), conduction and convection heat transfers by air can be safely neglected. Figure \ref{Fig:2}(a) shows a cross-sectional view of the experimental setup and describes the heat flow within the system. During the experiment, the feedback control of the input power (denoted as $Q_{\text{in}}$) to the heater can maintain the temperature of the emitter at the desired value. $Q_{\text{in}}$ is then divided into $Q_{\text{e-r}}$ (i.e., the radiative heat transfer between the emitter and the receiver) and $Q_{\text{loss}}$ that includes the background radiation as well as the parasitic conduction from the heater to the vacuum chamber through the three-axis nanopositioner. Considering that $Q_{\text{e-r}}$ is the summation of far-field ($Q_{\text{e-r,far}}$) and near-field ($Q_{\text{e-r,near}}$) contributions, $Q_{\text{in}}$ can be expressed as $Q_{\text{in}}= (Q_{\text{e-r,near}}+Q_{\text{e-r,far}} )+Q_{\text{loss}}$. If the temperature of the vacuum chamber is maintained at a constant temperature (e.g., room temperature), $Q_{\text{ref}}=Q_{\text{e-r,far}}+Q_{\text{loss}}$ can also be considered as a constant while reducing the gap between the emitter and the receiver. In our experimental condition, it was confirmed that $Q_{\text{ref}}$ can be regarded as constant within 5 minutes (see Supplementary Fig.\ S7) and the standard deviation of the $Q_{\text{ref}}$ is considered in error estimation of the obtained data. Therefore, one cycle of the data acquisition from the vacuum gap of 2200 nm to the vacuum gap where the first local contact between the emitter and the receiver is detected was performed within 5 minutes. To measure the radiative heat flux between the emitter and the receiver with respect to the vacuum gap, we firstly measured $Q_{\text{in}} \approx  Q_{\text{ref}}$ at the vacuum gap of 2200 nm where $Q_{\text{e-r,near}}$ is negligible compared to $Q_{\text{e-r, near}}$ at the vacuum gap of 380 nm (i.e., the smallest vacuum gap achieved). As the vacuum gap is decreasing, $Q_{\text{in}}$ required to maintain the emitter temperature constant is increased due to the contribution of the evanescent mode, i.e., $Q_{\text{in}}(d)=Q_{\text{e-r,near}}(d)+Q_{\text{ref}}$. In this way, we could estimate $Q_{\text{e-r}}$ by adding a calculated $Q_{\text{e-r,far}}$ to measured $Q_{\text{e-r,near}}$. In the upper panel of Fig.\ \ref{Fig:2}(b), it can be clearly seen that the estimated $Q_{\text{e-r}}$ increases while the temperature of the heater is maintained within $\pm0.1$ K of the designated value. The corresponding vacuum gaps derived from the measured capacitances between capacitor electrodes are shown in the lower panel of Fig.\ \ref{Fig:2}(b) and the achieved parallelism (defined as the difference between $d_{1}$ and $d_{4}$) is 31.7 nm when the average vacuum gap is 380 nm, which corresponds to the tilting angle of $2.3 \times 10^{-6}$ rad. Note that compared to the previous work \cite{lim2018tailoring}, this differential-input-power method is much more straightforward than the heat flux estimation from the temperature differences between two thermistors based on the calibration result. Furthermore, the fabrication process for the receiver part of the microdevice is significantly simplified, because there is no need to integrate thermistors and a calibration heater. 

%%%%%%%%%%
\begin{figure}[!t]
\centering\includegraphics[width=0.6\textwidth]{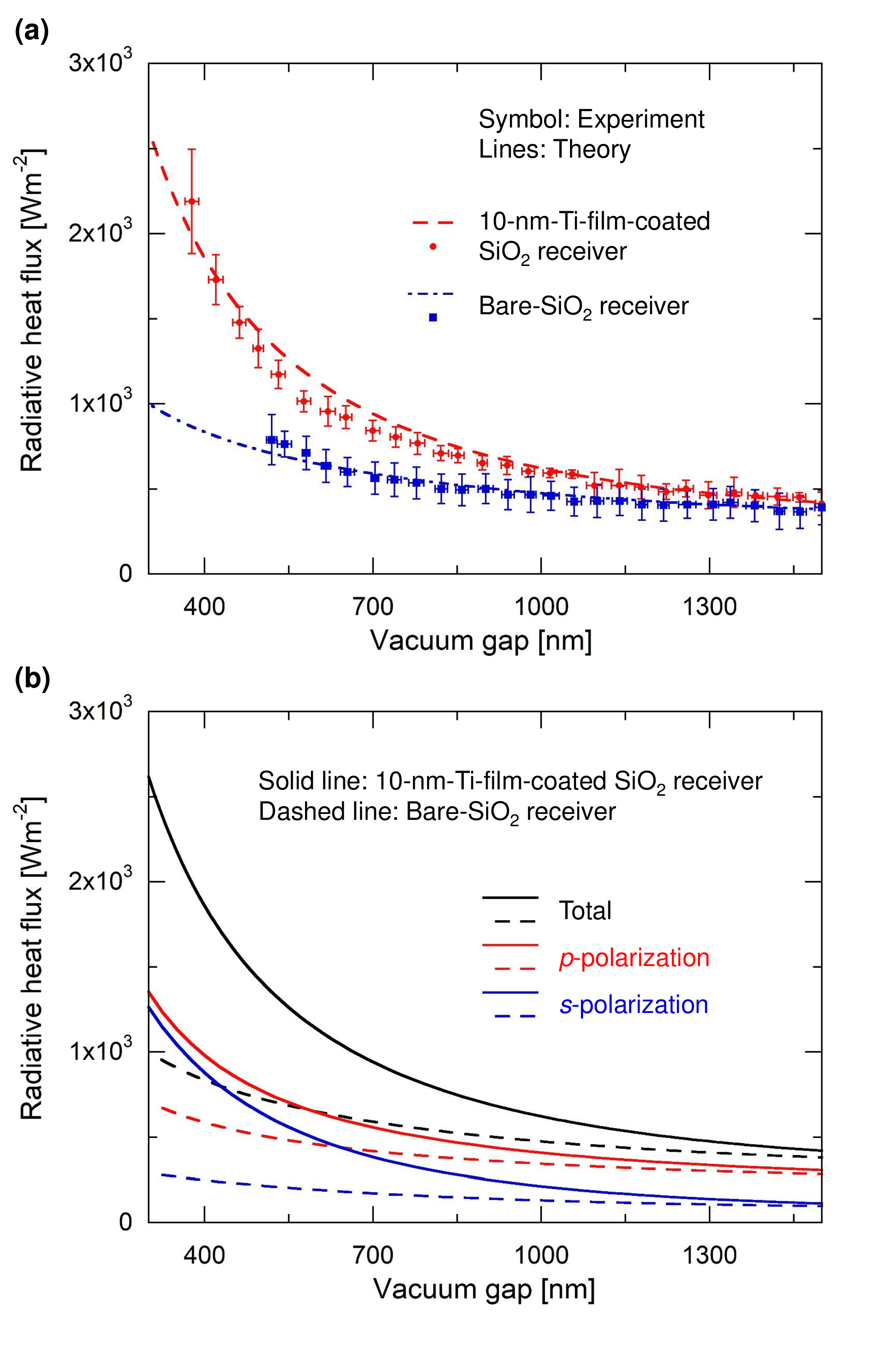}
\caption{\textbf{(a)} Measured near-field radiative heat flux between doped-Si emitters and SiO$_2$ with and without 10-nm Ti film. \textbf{(b)} Theoretically predicted value of near-field radiative heat flux for both configurations. Contributions of $p$- and $s$- polarization are also plotted. The emitter temperature is set as 430 K while maintaining receiver temperature at 300 K.}
\label{Fig:3}
\end{figure}
%%%%%%%

The measured radiative heat flux is plotted with respect to the vacuum gap ($d$) in Fig.\ \ref{Fig:3}(a). The doped-Si emitters for both configurations are set to 430 K while the receivers, which are the SiO$_2$ substrates with or without 10-nm Ti film overlaid, are maintained at 300 K. The data plotted in Fig.\ \ref{Fig:3}(a) are the values averaged from 10 independent experiments for each case. Although the minimum vacuum gap achieved for the case of bare SiO$_2$ is slightly larger than that for the case with 10-nm-thick Ti film, it can be seen that the measured values excellently agree with the theoretical predictions for both configurations. At $d=380$ nm, the measured radiative heat flux for the case with 10-nm-thick Ti film is around 2190 W/m$^2$, which is greater by 1860 W/m$^2$ than that at $d=2200$ nm, i.e., $Q_{\text{e-r}}(d=380 \text{ nm})-Q_{\text{e-r}}(d=2200 \text{ nm})=1860$ W/m$^2$. The value of $[Q_{\text{e-r}}(d=380 \text{ nm})-Q_{\text{e-r}}(d=2200 \text{ nm})]$ achieved with 10-nm Ti film is almost 3.5 times greater than the predicted enhancement for the case with the bare SiO$_2$ receiver. Such a considerable near-field enhancement is attributed to stronger coupling of SPPs confined at vacuum/doped-Si and vacuum/Ti-film/SiO$_2$ interfaces. As can be noted in Fig.\ \ref{Fig:3}(b), however, this enhancement not only results from the increase of heat flux via $p$-polarization (i.e., plasmonic contribution), but also that via $s$-polarization. Thus, more detailed analysis is needed.

The net near-field radiation between the emitter and the receiver can be expressed as \cite{polder1971theory, mulet2002enhanced, joulain2005surface,basu2010near}:
\begin{equation}
q=\int_0^{\infty} d\omega\ (q^{p}_{\omega} + q^{s}_{\omega}) =  \int_0^{\infty} d\omega\ \int_0^{\infty} \frac{\Theta(\omega, T_1)-\Theta(\omega, T_2)}{\pi^2} \times \left[ Z^{p}_{\beta, \omega}(\beta, \omega) + Z^{s}_{\beta, \omega}(\beta, \omega) \right ] d\beta,
\end{equation}
where $\omega$ stands for the angular frequency and $\beta$ is the parallel component of the wavevector. For this study, the temperature of the emitter, $T_1$ is 430 K and that of the receiver, $T_2$ is 300 K. Also, $\Theta(\omega,T_{i})= \frac{\hbar \omega}{\exp \lbrace \hbar \omega/(k_B T_i) \rbrace-1}$ is the mean energy of the Planck oscillator, where $\hbar$ is the Planck constant divided by $2\pi$ and $k_B$ is the Boltzmann constant. In order to elucidate the enhancement mechanism through coupling of surface polaritons, the effect of the Planck distribution $\Theta(\omega,T_{i})$ is sometimes excluded \cite{basu2010near, karalis2016squeezing, yang2018observing,shi2019colossal}, such that the analysis could be conducted based solely on the exchange function $Z_{\beta, \omega}(\beta,\omega)$. The exchange function can be expressed for both cases with or without 10-nm-thick Ti film as \cite{basu2010near}:
\begin{equation} \begin{aligned}
&Z_{\beta, \omega, prop}^{p,s}(\beta, \omega) = \frac{\beta (1-|r_{01}^{p,s}|^2)(1-|r_{02}^{p,s}|^2)}{4|1-r_{01}^{p,s} r_{02}^{p,s} e^{i2k_{0z}d}|^2} \\ 
& Z_{\beta, \omega, evan}^{p,s}(\beta, \omega) =\frac{\beta \text{Im}(r_{01}^{p,s})\text{Im}(r_{02}^{p,s}) e^{-2 \text{Im}(k_{0z})d}}{|1-r_{01}^{p,s} r_{02}^{p,s} e^{i2k_{0z}d} |^2},
\end{aligned} \end{equation}
where the expression for propagating and evanescent waves can be used when $\beta< \omega/c_0$ and $\beta> \omega/c_0$, respectively. In above equations, $k_{0z}$ is the normal component of wavevector in vacuum and Im() takes the imaginary part of a complex value. $r_{01}^{p,s}$ stands for the reflection coefficient at the vacuum/doped-Si interface and $r_{02}^{p,s}$ can be the reflection coefficient at the vacuum/SiO$_2$ interface or the modified reflection coefficient for the vacuum/Ti-film/SiO$_2$ multilayered structure, obtained using Airy's formula \cite{yeh1988optical, biehs2007thermal}. The optical property of Ti film was obtained from \cite{ordal1988optical} including the electron-boundary scattering effect \cite{ijaz1978electron, ding2015thickness} and that of doped Si was taken from \cite{basu2010near} by assuming complete ionization at high temperature. The dielectric function of SiO$_2$ reported in \cite{palik1998handbook} was employed.

%%%%%%%%%%
\begin{figure}[!b]
\centering\includegraphics[width=0.6\textwidth]{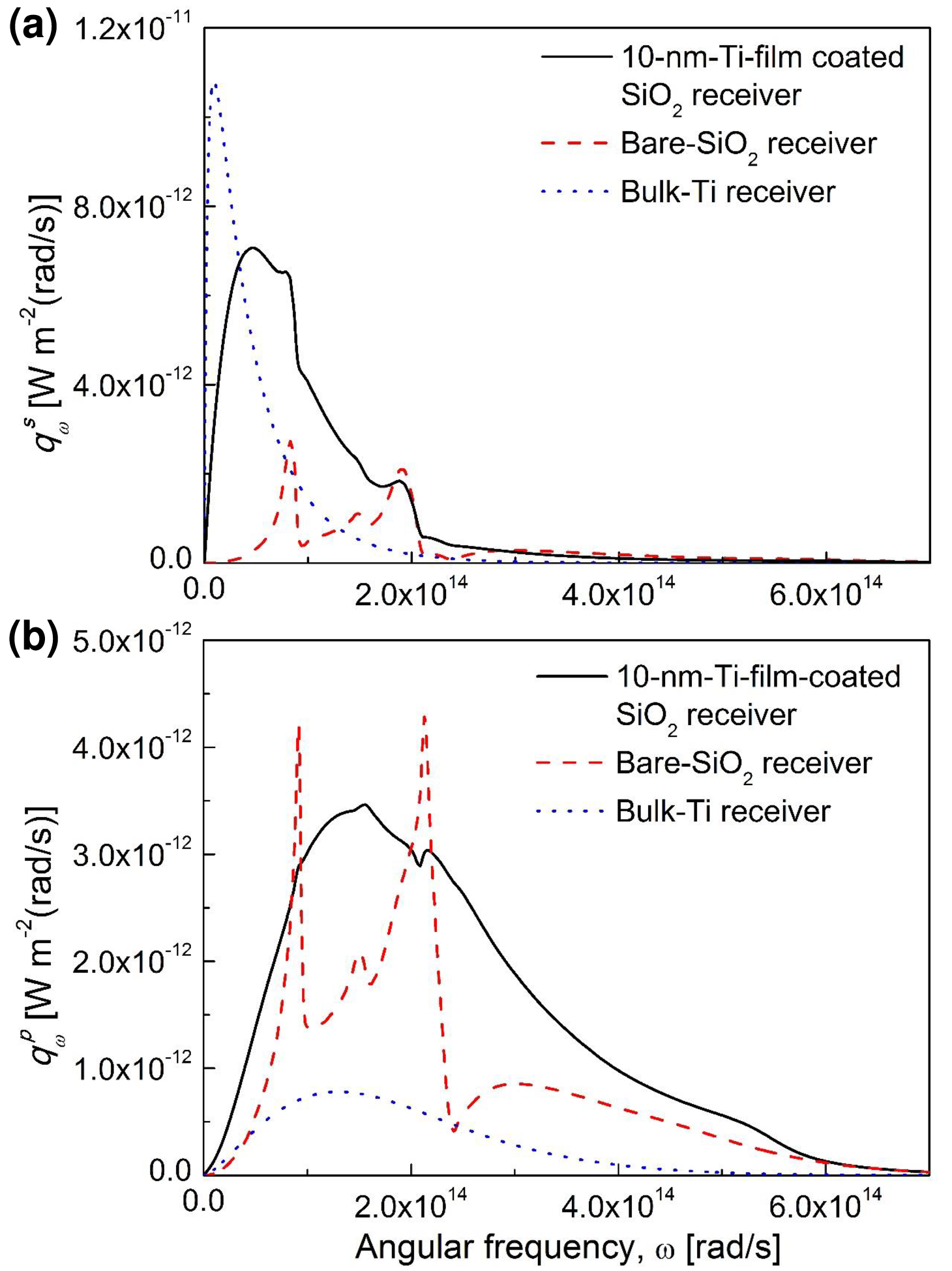}
\caption{The spectral heat flux, $q^{\gamma}_{\omega}$ for \textbf{(a)} $s$-polarization and \textbf{(b)} $p$-polarization at $d=400$ nm. The emitter is set as doped-Si at 430 K while three different structures (i.e., 10-nm-Ti-film-coated SiO$_2$, bare SiO$_2$, and bulk Ti) are employed as a receiver at 300 K.}
\label{Fig:4}
\end{figure}
%%%%%%%

In Figs.\ \ref{Fig:4}(a)-\ref{Fig:4}(b), the calculated spectral radiative heat flux for each polarization is depicted when $d= 400$ nm. For both polarizations, it can be clearly seen that 10-nm-thick-Ti-film-coated SiO$_2$ leads to the greatest heat transfer rate among three cases. In other words, we can see so-called `thin-film effect' \cite{biehs2007thermal2, francoeur2008near} in both polarizations for the configuration with 10-nm-thick-Ti-film-coated SiO$_2$. 

Let us discuss the enhancement of heat flux via $s$-polarization first. It is well-known that the heat transfer via $s$-polarization is dominant for the near-field radiation between metals \cite{chapuis2008effects}, while that through $p$-polarization is dominant in heat transfer between a metal and a polar material or between polar materials \cite{biehs2007thermal,fu2009near}. Due to high doping concentration of doped Si, the near-field radiative heat transfer between semi-infinite doped Si and semi-infinite (i.e., bulk) Ti is dominated by $s$-polarization. Thus, by placing thin Ti film, heat transfer through $s$-polarization can be increased as can be seen in Fig.\ \ref{Fig:4}(a). Comparing to the bulk Ti case, the peak in spectral heat flux for the case of 10-nm-thick Ti film is broadened and shifted to a higher frequency due to the increased electron-boundary scattering \cite{lim2018tailoring}. Because the spectral near-field energy density above the thin-metal-film-coated surface is maximized at smaller thickness with increasing frequency \cite{biehs2007thermal2}, 10-nm-thick Ti film can even show larger spectral heat flux than bulk Ti at higher frequency, leading to an increase in total heat transfer. Consequently, the Ti-film-coated surface can result in a significant near-field heat transfer in $s$-polarization compared to that between bulk doped Si and bulk SiO$_2$, and it can even exceed that between bulk doped Si and bulk Ti (e.g., when $d>180$ nm, for the configuration defined here).

\begin{figure}[!t]
\centering\includegraphics[width=0.95\textwidth]{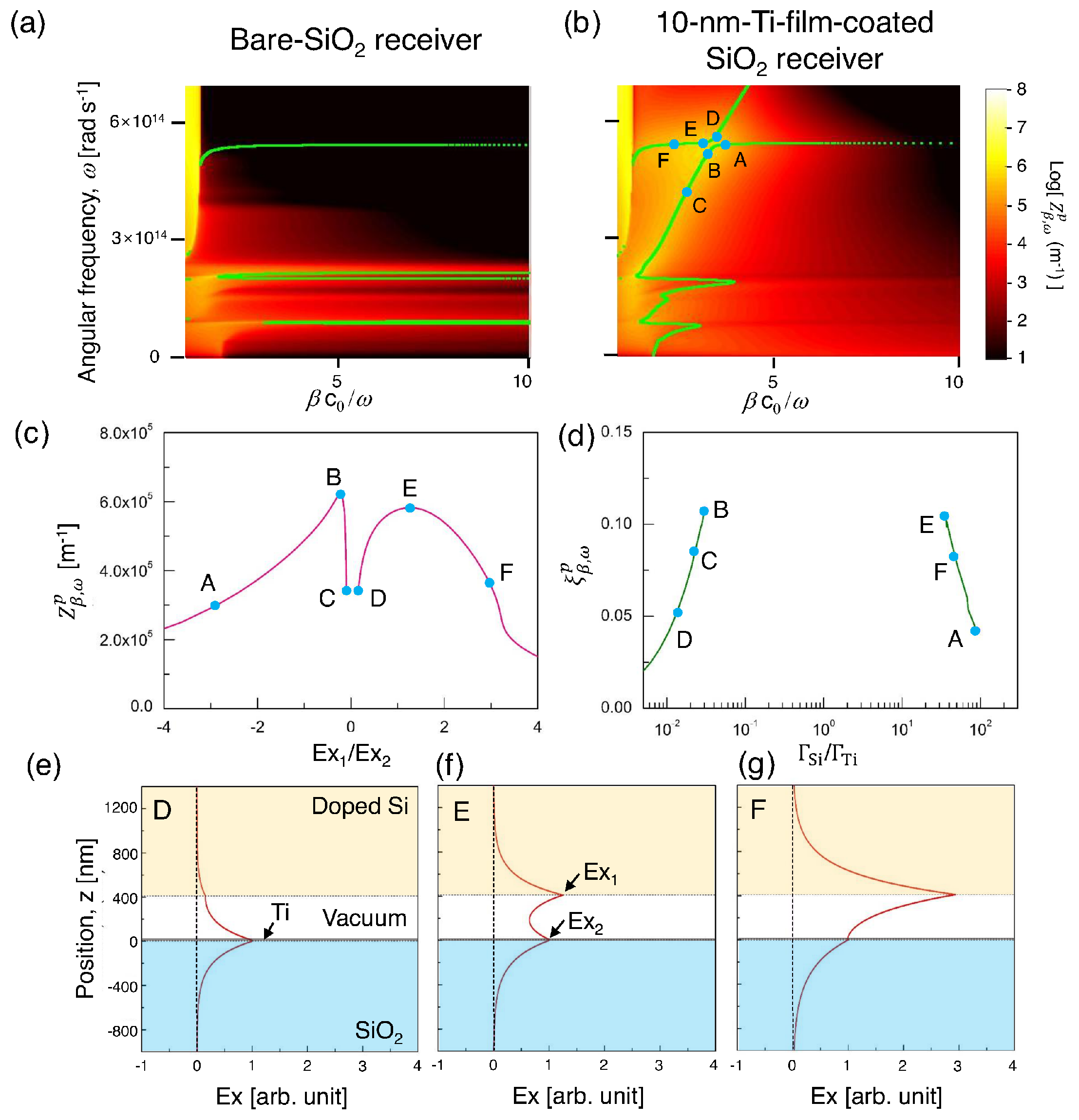}
\caption{Contour plot of exchange function $Z^{p}_{\beta, \omega}$ with respect to the normalized parallel wavevector and the angular frequency. The green-curves depicting SPP dispersion relations are overlaid. \textbf{(a)} Receiver of bare SiO$_2$. \textbf{(b)} Receiver of 10-nm-Ti-film-coated SiO$_2$. \textbf{(c)} Exchange function $Z^{p}_{\beta, \omega}$ plotted with respect to the longitudinal electric  field ratio (i.e., $E_{x1}/E_{x2}$). \textbf{(d)} Transmissivity with respect to the loss rate (i.e., $\Gamma_\text{Si}/\Gamma_\text{Ti}$) in coupled mode theory.  \textbf{(e)}, \textbf{(f)}, and \textbf{(g)} The longitudinal electric field profiles at the points D, E, and F. }
\label{Fig:5}
\end{figure}

For the near-field radiative heat flux via $p$-polarization, the quasi-monochromatic spectral radiative heat flux is observed for the bare-SiO$_2$ receiver [see Fig.\ \ref{Fig:4}(b)]. This is because the coupling of SPPs supported at the vacuum/doped Si interface and SPhPs supported at the vacuum/SiO$_2$ occurs only in a narrow frequency range. On the other hand, by coating the SiO$_2$ surface with a 10-nm-thick Ti film, a broad spectral enhancement can be achieved, leading to increase in a total amount of heat transfer by $p$-polarization. This mechanism of heat transfer enhancement is clearly revealed with the exchange function $Z^{p}_{\beta, \omega}$ plotted with respect to the normalized parallel wavevector and the angular frequency [refer to Figs.\ \ref{Fig:5}(a)-\ref{Fig:5}(b)]. The dispersion curves for the surface waves bounded in the entire structure can be obtained by neglecting losses of materials \cite{yeh2008essence, ordonez2013anomalous, ordonez2014thermal} and shown as green-colored-curves in the figure. 

%%%%%%%%%%
%%%%%%%

In Fig.\ \ref{Fig:5}(a), $Z^{p}_{\beta, \omega}$ has greater values at the frequencies $\sim 2.1\times 10^{14}$ rad/s and $\sim 9.1\times 10^{13}$, which corresponds to the SPhP resonant frequencies of vacuum/SiO$_2$ interface \cite{hu2008near, song2016radiative}. This increase in $Z^{p}_{\beta, \omega}$ results from the coupling between SPhPs of vacuum/SiO$_2$ interface and photon-like SPP mode of vacuum/doped-Si interface (i.e., near vacuum light line). Although a dispersion curve still appears near the SPP resonant frequency for the vacuum/doped Si interface ( $\sim$$5.4\times 10^{14}$), it cannot lead to an enhancement in $Z^{p}_{\beta, \omega}$, given that SPhP resonant mode for vacuum/SiO$_2$ interface does not exist in such a high frequency. On the other hand, when a thin Ti film is coated on the surface of SiO$_2$, the SPP condition of vacuum/receiver interface is modified such that it can be excited over broad spectral range. As a result, the coupled SPP-SPhPs existing at the vacuum/Ti-film/SiO$_2$ structure can now be interacted with the SPPs of vacuum/doped Si interface [see Fig.\ \ref{Fig:4}(b)]. Further, SPP dispersion curves are splited when two SPP dispersion curves of vacuum/doped Si and vacuum/receiver crosses each other near points B and E. This split of SPP dispersion curves has been reported in several previous studies \cite{francoeur2010local, iizuka2018significant, karalis2016squeezing} and is responsible for strong enhancement of the heat transfer as also can be seen in Fig.\ \ref{Fig:5}(b). 

To quantify the coupling strength of SPPs, we propose a dimensionless parameter based on the electric field amplitude at the vacuum interfaces. For detailed analysis, six points, indicated as A-F in Fig.\ \ref{Fig:5}(b), are selected along the polariton dispersion curves. When the longitudinal electric field component, $E_x$ is considered, the ratio of its amplitude at the vacuum-emitter interface (i.e., $E_{x1}$) and the vacuum-receiver interface (i.e., $E_{x2}$) can be calculated along the surface-wave dispersion curves. In Fig.\ \ref{Fig:5}(c), $Z^{p}_{\beta, \omega}$ is plotted with respect to the longitudinal electric field ratio (i.e., $E_{x1}/E_{x2}$) along the dispersion curves. It can be readily seen that $E_{x1}/E_{x2}$ curves are divided into two branches, which corresponds to each of SPP dispersion curves in Fig.\ \ref{Fig:5}(b). The sign of $E_{x1}/E_{x2}$ reveals that the SPP dispersion curve including points A-B-C corresponds to the antisymmetric mode of excitation, whereas the branch including points D-E-F is the symmetric mode. More importantly, the magnitude of $E_{x1}/E_{x2}$ can provide additional information about how effectively the surface wave at the emitter-vacuum interface and that at the receiver-vacuum interface are coupled together. If the surface wave at the one of the vacuum interfaces dominates, then the resulting $E_{x1}/E_{x2}$ value will tend toward either $|E_{x1}/E_{x2}| \ll 1$ or $|E_{x1}/E_{x2}| \gg 1$. In Fig.\ \ref{Fig:5}(e), for instance, the $E_x$ field distribution at point D is depicted where the field is mostly bounded at the vacuum/receiver interface, resulting in $E_{x1}/E_{x2}\approx 0.16$. In contrast, for point F, the field is mainly bounded at the vacuum/emitter interface [see Fig.\ \ref{Fig:5}(g)], and the corresponding $E_{x1}/E_{x2}$ value is 3.0. Comparison of Figs.\ \ref{Fig:5}(a) and \ref{Fig:5}(b) reveals that the dispersion curve where point F locates in Fig.\ \ref{Fig:5}(b) is actually originated from the SPP dispersion of the vacuum/doped-Si interface, while that of point D is from the SPP dispersion of the vacuum/receiver interface. Thus, it can be inferred that too low or too high values of $|E_{x1}/E_{x2}|$ indicate that the SPPs of the vacuum/emitter and the vacuum/receiver interfaces are unbalanced (i.e., weakly coupled). At point E where the maximum $Z^{p}_{\beta, \omega}$ occurs, however, the SPPs from each interface are strongly coupled, such that magnitude of evanescent waves at the emitter and the receiver are in similar range (i.e., $E_{x1}/E_{x2}=1.3 \approx 1$). For the antisymmetric branch [i.e., points A-B-C in Fig.\ \ref{Fig:5}(c)], the similar behavior of $|E_{x1}/E_{x2}|$ can be observed, but the maximum $Z^{p}_{\beta, \omega}$ occurs when $|E_{x1}/E_{x2}|=0.22$. 

In the previous studies \cite{karalis2015temporal, karalis2016squeezing, lin2017application}, the maximum thermal transmissivity, $\xi^{p}_{\beta, \omega}$ ($=Z^{p}_{\beta, \omega} / \beta$) is often estimated from the impedance matching condition derived from coupled mode theory. Figure \ref{Fig:5}(d) shows the transmissivity, $\xi^{p}_{\beta, \omega}$ with respect to the ratio of the loss rates (i.e., $\Gamma_\text{Si}$ and $\Gamma_\text{Ti}$; refer to \cite{karalis2015temporal, lin2017application2} for details) obtained by activating loss of each layer. As predicted, $\xi^{p}_{\beta, \omega}$ increases as two loss rates become closer (i.e., $\Gamma_\text{Si}/\Gamma_\text{Ti}$ approaches to 1). Because the vacuum gap is 400 nm, the maximum transmissivity is bound to 0.12, which can be obtained at the points B and E [see Fig.\ \ref{Fig:5}(d)]. Although it is not shown here, the maximum transmissivity can reach 1.0 when $d=100$ nm, at which $\Gamma_\text{Si}/\Gamma_\text{Ti} \approx 1$. Similar to $E_{x1}/E_{x2}$,  the values of $\Gamma_\text{Si}/\Gamma_\text{Ti}$ are divided into two branches. Because the dispersion curve including points B, C, and D originates from the SPP dispersion of the vacuum/receiver interface, $\Gamma_\text{Ti}$ (i.e., loss to the Ti film) is larger than $\Gamma_\text{Si}$ (i.e., loss to doped Si). Similarly, for dispersion curves including points E, F, and A, $\Gamma_\text{Si}$ has larger value than $\Gamma_\text{Ti}$, because the dispersion curve including points E,F, and A stems from the SPP dispersion curves for vacuum/doped-Si interface. Interestingly, the six points A-F are divided differently for $E_{x1}/E_{x2}$ and $\Gamma_\text{Si}/\Gamma_\text{Ti}$, providing complemental information. As a matter of fact, $E_{x1}/E_{x2}$ can provide the information on the mode profile of plasmonic resonances, but $\Gamma_\text{Si}/\Gamma_\text{Ti}$ can give the information on the decay rate to each layer for a given resonant mode. These two parameters can be used together to predict the coupling strength of the SPPs and to analyze the optimal configuration maximizing the heat transfer. Finally, it is worthwhile to mention that although strong coupling is observed at points B and E, because of Planck distribution, it cannot result in prominent enhancement in the spectral heat flux shown in Fig.\ \ref{Fig:4}(b). Nevertheless, the relationship between $Z^{p}_{\beta, \omega}$ and the field ratio along the dispersion curve is still valid at a lower frequency regime and significant enhancement is expected when the doping concentration of the emitter becomes low.

In summary, we have suggested a means to strongly increase the near-field radiative heat transfer between doped Si and SiO$_2$ media by coating SiO$_2$ substrate (i.e., polar material) with a thin Ti film and successfully demonstrated increased near-field heat transfer using a custom-built MEMS-integrated platform. While decreasing vacuum gap from 2200 nm to 380 nm, the enhancement in near-field radiative heat transfer between doped-Si emitter and 10-nm-thick Ti-film-coated SiO$_2$ receiver is measured to be 1860 W/m$^2$ which is 3.5 times larger value than that for the case of doped-Si emitter and bare SiO$_2$ receiver. It was revealed that this thin-metal film can enhance the heat transfer both via $p$- and $s$- polarizations. In particular, the heat transfer enhanced via $p$-polarization results from the coupling of SPPs from vacuum-emitter and vacuum-receiver interfaces in a broad spectral range. This enhancement is predicted with the proposed dimensionless parameter, which is the ratio of the longitudinal electric field at vacuum/emitter and vacuum/receiver interfaces. Considering that thin metal film is compatible with the engineering applications of near-field radiative heat transfer such as a Schottky-junction based near-field TPV system \cite{st2017hot, vongsoasup2018effects}, the results obtained in this study will guide the future development of the high-throughput near-field devices.

\begin{acknowledgments}
This research was supported by the Basic Science Research Program (NRF-2017R1A2B2011192, NRF-2018R1A6A3A01012563 and NRF-2019R1A2C2003605) through the National Research Foundation of Korea (NRF) funded by Ministry of Science and ICT.  
\end{acknowledgments}

\bibliography{Lim_Bib}

\end{document}